# Electron paramagnetic resonance of n-type semiconductors for applications in 3D thermometry

*Darshan Chalise [1,2\*], David G. Cahill[1,2,3]*

*[1]Department of Physics, University of Illinois at Urbana-Champaign, Urbana, IL, 61801, USA*

*[2]Materials Research Laboratory, University of Illinois at Urbana-Champaign, Urbana, IL, 61801, USA*

*[3]Materials Science and Engineering, University of Illinois at Urbana-Champaign, Urbana, IL, 61801, USA*

\*Corresponding Author – darshan2@illinois.edu

## Abstract

While there are several 2D thermometry techniques that provide excellent spatial, temporal and time resolution, there is a lack of 3D thermometry techniques that work for a wide range of materials and offer good resolution in time, space and temperature. X-ray diffraction (XRD) and nuclear magnetic resonance (NMR) imaging can provide 3D temperature information. However, XRD is typically limited to crystalline materials while NMR is largely limited to liquids where the resonance lines are sufficiently narrow. We investigate electron paramagnetic resonance (EPR) of n-type silicon and germanium as a possible means of 3D thermometry. While in germanium the EPR linewidths are too broad for thermometry, EPR linewidths in silicon are reasonably narrow and exhibit a strong temperature dependence. The temperature dependence of the spin-lattice relaxation rate ($1/T_1$) of conduction electrons in n-type Si have been extensively studied for low dopant concentrations and follows a $T^3$ law due to phonon broadening. For heavily doped Si, which is desirable for good signal to noise ratio (SNR) for application in thermometry, impurity scattering is expected to decrease the temperature dependence of $1/T_1$. Our results show that, in heavily doped n-type Si, spin-lattice relaxation induced by impurity scattering does not drastically decrease the temperature dependence of EPR linewidths. In P-doped Si with donor concentration of $7 \times 10^{18}$ /cm$^3$, the EPR linewidth has a $T^{5/2}$ temperature dependence; the temperature dependence decreases to $T^{3/2}$ when the donor concentration is $7 \times 10^{19}$ /cm$^3$. While the temperature dependence of linewidth decreases for heavier doping, EPR linewidth is still a sensitive thermometer. We define a figure of merit for SNR for thermometry from EPR linewidths of n-type Si and observe that increasing the doping results in a better SNR for thermometry. Using effective medium theory, we show that EPR linewidth can be a sensitive thermometer for application in 3D thermometry with systems embedding microparticles of heavily doped n-type Si.

# I. INTRODUCTION

While sensitive and spatially well resolved invasive [1] and non-invasive [2] 2-dimensional (2D) thermometry techniques have been established, there is a general lack non-invasive 3-dimensional (3D) thermometry. This is especially true in systems where there is no optical access. 3D thermometry in optically opaque systems is, however, important in biology [3] and engineering [4].

Imaging of systems opaque to visible light can be performed using electromagnetic waves of much higher (x-ray) or lower frequencies (radio frequencies or microwave frequencies) [5]. 3D X-ray imaging is possible in both absorption [6] and diffraction mode [7]. However, X-ray absorption is not strongly temperature dependent and X-ray diffraction requires crystalline materials [4]. Even for crystalline materials, XRD requires sufficiently high energy x-rays for 3D thermometry [4] .

Different nuclear magnetic resonance (NMR) processes depend on temperature [3], and therefore, nuclear magnetic resonance imaging, i.e., (N)MRI, can be used to obtain spatially resolved 3D temperature maps. In biological systems, (N)MRI has been used to obtain temperature maps using spin-density contrast [8], spin-lattice relaxation time ($T_1$) contrast [9], spin-spin relaxation time ($T_2$) contrast [10], [11] , chemical shift contrast [12,13] and diffusion contrast [11] [14] . While spin density and the relaxation time contrasts are relatively less sensitive to temperature, diffusion and chemical shift contrast suffer from motional artifacts [3]. Additionally, all these techniques depend on the nature of water for the temperature dependence of the NMR signal [3], and therefore, cannot be applied in systems not containing water.

In our previous work, we showed temperature dependent diffusion results in temperature dependent motional narrowing of NMR lines in the presence of superparamagnetic iron oxide nanoparticles (SPIONs). The subsequent spin-spin relaxation time ($T_2$) is a thermometer in fluids in the presence of SPIONs [15]. The method, however, is restricted to fluids where the timescale of diffusion is fast enough to even out the field inhomogeneities induced by the SPIONs. The field inhomogeneities induced by the SPIONs (the magnetization of the SPIONs) is not strongly temperature dependent. The method is, therefore, limited to fluids where the temperature dependence of diffusion is strong.

In general, NMR is limited to systems where the nucleus of interest is has a fast enough diffusion to give an observable NMR spectrum. This makes NMR thermometry not applicable in most solids.

3D imaging using magnetic resonance, however, is not limited to nuclear spin resonances. Electron paramagnetic resonance (EPR) could also be used for 3D imaging of systems with paramagnetic electrons or free radicals [16]. Temperature dependent EPR signal has also suggested for possible thermometry [17]. Therefore, EPR Imaging (EPRI) provides an opportunity for 3D thermometry in systems where x-ray or NMR cannot obtain the temperature information.

In the first paper reporting EPR in semiconductors, Kittel and Brattain [18] noted that the EPR linewidth of conduction electrons in heavily doped n-type silicon is highly sensitive to temperature at elevated temperatures (>100 K). Following their work, theories by Elliot [19] and Yafet [20] and further improved by others [21–23] showed that phonon broadening results in a $\sim T^3$ temperature dependence of EPR linewidth in Si. This shows, EPR linewidth in n-type Si can be a sensitive 3D thermometer.

For heavily doped Si, which is preferred for a higher signal to noise ratio (SNR) in thermometry, impurity scattering is expected to impact the line broadening of EPR spectrum [24–26]. There is a need for a systematic study on how the impurity impacts the temperature dependence of EPR linewidths at higher temperatures.

In this paper, we systematically investigate the temperature dependence of EPR linewidth of n-type Si at different carrier densities to identify the optimal doping level for application in thermometry. We also study the EPR linewidths in n-type Ge to investigate if Ge provides a better opportunity for 3D thermometry.

Finally, we apply effective medium theory to estimate the penetration depths of microwave at different doping levels and frequencies.

The results of this study should allow selecting the most appropriate doping level and the type of semiconductor for strong temperature dependence as well as good signal to noise ratio (SNR) required for 3D thermometry.

## II. EXPERIMENTAL DETAILS

P-doped Si with different carrier concentrations were purchased from Sil'Tronix Silicon Technologies and As-doped and Sb-doped Ge were purchased from El-Cat Inc. The room temperature sheet resistance and resistivity of the wafers was measured by using a 4-point probe using a constant current source and Keithly 2000 as a DC voltmeter. Table 1 summarizes the measured room temperature resistivities and corresponding carrier concentrations for each sample.

Table 1. Resistivities and carrier densities for studied samples.

| Sample no. | Type | Resistivity (m$\Omega$.cm) | Carrier concentration, $n_c$ (cm$^{-3}$) |
|---|---|---|---|
| A | P-doped Si | 1.01 | $7 \times 10^{19}$ [27] |
| B | P-doped Si | 8.3 | $7 \times 10^{18}$ [28] |
| C | P-doped Si | 105 | $7 \times 10^{16}$ [27] |
| D | P-doped Si | 4300 | $9 \times 10^{14}$ [27] |
| E | P-doped Si | 528000 | $7 \times 10^{12}$ [28] |
| F | As-doped Ge | 25 | $5 \times 10^{17}$ [29] |
| G | Sb-doped Ge | 7.9 | $1.5 \times 10^{18}$ [29] |

Temperature controlled EPR measurements from 100 K – 290 K were carried out on a Varian EMXPlus spectrometer at a central frequency of ~ 9.4 GHz (X-band). Central field was swept with a modulation frequency of 100 kHz. For measurements on Si, the modulation amplitude for most measurements were set at 1 G but was reduced if the EPR linewidths were comparable to or smaller than 1 G. For Ge, the EPR lines were very broad (~40 G) for all temperatures, and therefore, a modulation amplitude of 10 G was used to maximize the signal to noise ratio. To confirm the line-broadening was independent of the field used, additional measurements were also performed at an Elesyx1 spectrometer with central frequencies of 34 GHz (Q-band) and 3.5 GHz (S-band).

Samples not ground to powders showed a Dysonian lineshape [30,31] at the expected g-values of conduction electrons. This is expected for highly conductive samples where microwave penetration depths being smaller than the particle size [32] . The effect was more pronounced at

34 GHz as the penetration depth is reduced further. Therefore, the samples were ball-milled using alumina balls and jar and inspected with scanning electron microscopy to ensure the average particle sizes were smaller than the microwave penetration depth (10 μm for 1 mΩ.cm Si at 9.3 GHz), the Dysonian effect was removed, and the line shape could be fit well with Lorentzian function using EasySpin tool in Matlab.

Ball milling, however, resulted in an additional feature in the EPR spectrum, which has been attributed in n-type semiconductor to a paramagnetic state created due to dangling bonds formed by ball milling damage [33,34]. In silicon, it had been suggested that heating the sample in air at 800 °C anneal the damage to remove the feature [34]. We were able to remove this feature when the powdered samples were heated in a muffler oven at 800 °C.

The comparison of lineshapes before ball milling, after the ball milling and after heating the ball milled sample at 800 °C are included in Fig 1.

**a**

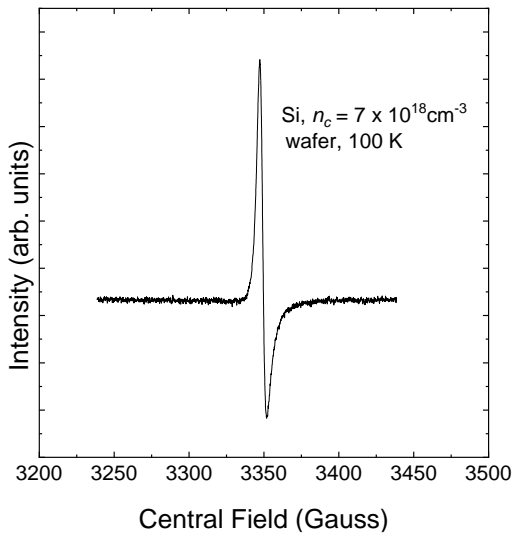

**b**

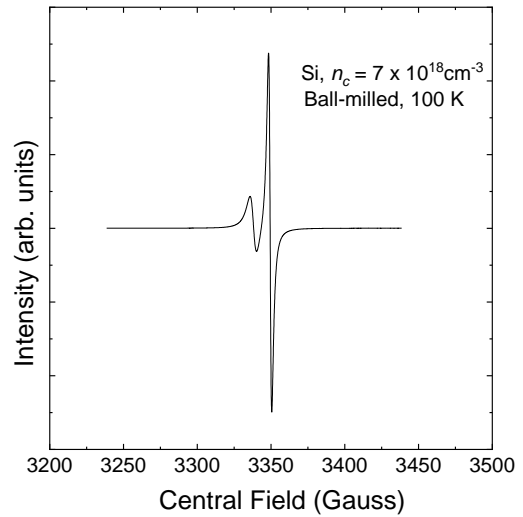

c

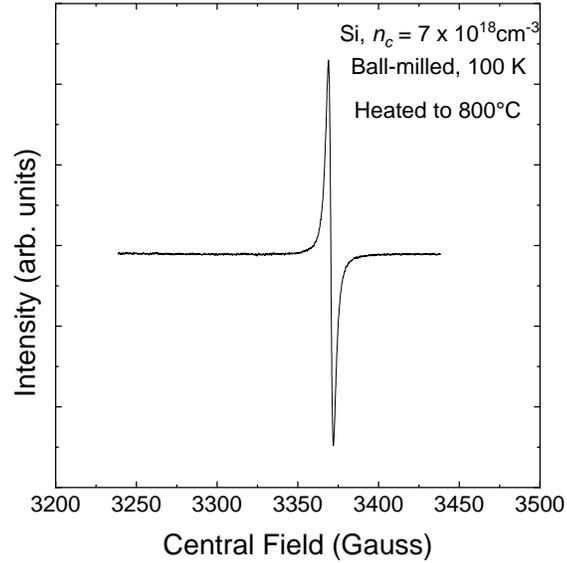

Fig 1. Comparison of EPR spectra in P-doped Si ($n_c = 7 \times 10^{18}$ cm$^{-3}$) at 100 K (a) before the sample is ball-milled, (b) after ball-milling but before heating at 800 °C and (c) after ball-milling and heating at 800 °C. The recorded spectra is a derivative spectra with a central field sweep of 200 G at a modulation amplitude of 1 G and a modulation frequency of 100 kHz. The peak to peak linewidth ($\Delta \nu_{p-p}$) represents the full-width at half-maximum of the absorption spectra. (a) shows a Dysonian lineshape as the particle size (300 $\mu$m thick wafer is smaller than the microwave penetration depth (~ 20 $\mu$m at 9.4 GHz). (b) shows the Dysonian effect is removed after ball-milling but ball-milling results in paramagnetic surface states that appear in the EPR spectra. (c) shows that annealing the sample at 800 °C removes the paramagnetic surface states.

## III. RESULTS

We were able to observe EPR of conduction electrons ($g \approx 1.998$ [35]) in P-doped Si with resistivities carrier densities $n_c = 7 \times 10^{19}$ cm$^{-3}$, $n_c = 7 \times 10^{18}$ cm$^{-3}$ and $n_c = 7 \times 10^{16}$ cm$^{-3}$ in the entire temperature range of our investigation i.e. 100 K – 290 K. We also observed EPR of conduction electrons ($g \approx 1.57$ [36,37]) in As-doped Ge ($n_c = 5\times 10^{17}$ cm$^{-3}$) from 50 K to 170 K. Above 170 K, the EPR linewidth was too broad for the spectra to be observable.

For P-doped Si with $n_c = 7\times 10^{12}$ cm$^{-3}$, we did not observe any EPR signal for expected values of conduction electrons (g = 1.998 [35]). We attribute this to poor SNR due to lack of conduction electrons contributing to spin resonance. Even for the Si with $n_c = 9\times 10^{14}$ cm$^{-3}$, the EPR signal was too weak to make conclusions on the dependence of the linewidth and integrated intensities on temperature. We also did not observe EPR in conduction electrons in Sb-doped Ge. The lack of observation in EPR in Sb-doped Ge has been attributed to strain broadening of the spectra which causes resonance to be too broad to be observed during the EPR experiment [38].

For conduction electrons in n-type semiconductors, the spin-spin relaxation ($1/T_2$) is suppressed due to motional narrowing of the conduction, and the observed linewidth of the EPR spectra is limited by spin-lattice relaxation ($1/T_1$) [19,39].

The temperature dependence of the spin-lattice relaxation rate of conduction electrons in semiconductors was first studied by Elliot predicting the spin-flip probabilities due to momentum scattering of the electrons with the lattice [19]. Yaffet also included the contribution of the modulation of spin-orbit coupling due to lattice vibration [20]. Elliot-Yaffet theory predicts a $T^{\frac{5}{2}}$ temperature dependence of the of the spin-lattice relaxation rate due to electron-phonon scattering for intravalley scattering of electrons by phonons [20,21]. The formalisms by Cheng et.al [21] and Park et.al [23] pointed out that at higher temperature intervalley scattering of electrons by phonons dominates the spin-lattice relaxation and the spin-lattice relaxation rate follows a $T^3$ power-law.

Fig 2 a. shows the temperature dependence of EPR linewidths of P-doped Si samples with $n_c = 7 \times 10^{16}$/cm$^3$, $n_c = 7 \times 10^{18}$/cm$^3$ and $n_c = 7 \times 10^{19}$/cm$^3$ as well as As-doped Ge with $n_c = 5 \times 10^{17}$/cm$^3$. The linewidths in Ge above 170 K are too broad, and the SNR is too low to make any conclusions about the temperature dependence.

The linewidths in Si are compared with experimental data from Lancaster [39] and Lepine [40] as well as the theory by Cheng et. al [21]. Our data for sample with $n_c = 7 \times 10^{16}$/cm$^3$ as well as the data from Lancaster and Lepine for similarly doped samples show that the predictions of EPR linewidth from Cheng et. al matches reasonably well for low doped samples and follows a $\sim T^3$ trend.

For heavily doped samples, the EPR linewidths are much broader than the phonon-broadened linewidth predicted by Cheng. For heavily doped samples, impurity scattering is expected to affect the spin-lattice relaxation rate [24–26]. Fig 2b includes results from subtracting the phonon-broadened linewidth. This remnant linewidth, which can be attributed to impurity scattering follows a temperature dependence that is only slightly smaller than $T^{3/2}$. This is in contrast to the formalism by Song et.al [26] where the temperature dependence of impurity scattering for heavily doped samples is expected to be $\sqrt{T}$. Therefore, results on heavily doped n-type Si show that the theory predicting the effect of impurity scattering on spin relaxation requires a revision.

For application in thermometry, the high temperature dependence of impurity scattering induced spin-lattice relaxation means that EPR linewidth is a highly sensitive thermometer for high doped silicon.

It is useful to define a sensitivity coefficient, $\xi_S^T$, which defines the percentage change in signal $S$ for 1% change in the absolute temperature $T$, $\xi_S^T = \frac{d \ln p}{d \ln T}$ [41] . For Si with carrier density, the temperature dependence of EPR linewidth is close to $T^3$ and $\xi_{\Delta\nu}^T \approx 3$.

For higher carrier densities, i.e., $n_c \approx 7 \times 10^{18}$ /cm$^3$ and $n_c \approx 7 \times 10^{19}$ /cm$^3$, the temperature dependence decreases to $\sim T^{5/2}$ and $\sim T^{3/2}$. Therefore, the respective sensitivities to temperature decrease to $\sim 2.5$ and $\sim 1.5$.

**a**

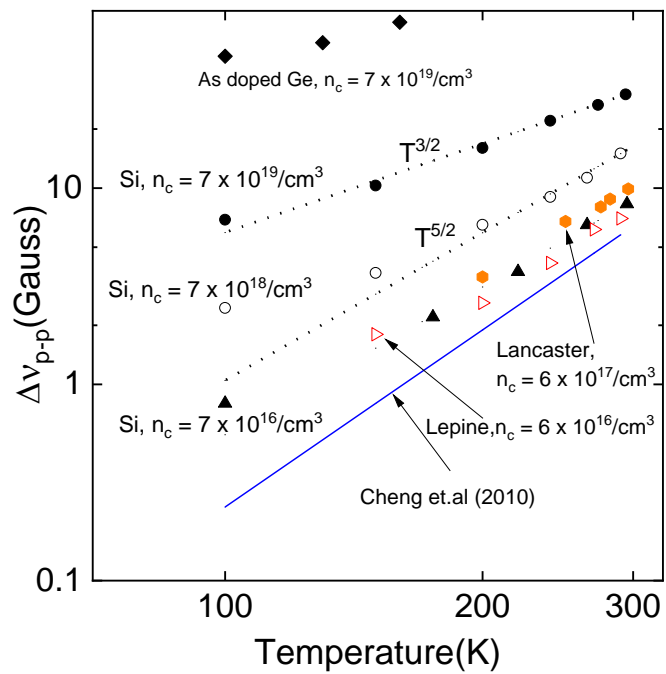

**b**

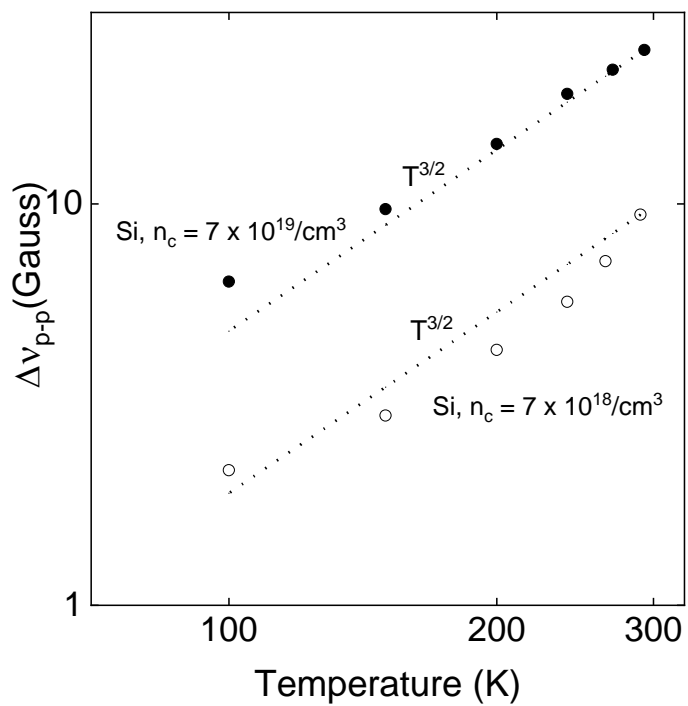

Fig 2. Temperature dependence of peak-to-peak EPR linewidth, $\Delta v_{p\text{-}p}$. (a) shows the experimental linewidths of P-doped Si with carrier densities $n_c = 7 \times 10^{19}$ cm$^{-3}$ (filled black circles), $n_c = 7 \times 10^{18}$ cm$^{-3}$ (open black circles), $n_c = 7 \times 10^{16}$ cm$^{-3}$ (filled black triangles) and As-doped Ge with $n_c = 5 \times 10^{17}$ cm$^{-3}$ (filled black diamonds). Experimental data by Lepine [40] (open red forward-triangles) and Lancaster [39] (filled orange hexagon) and the theoretical linewidth predicted by Cheng et.al [21] (blue solid line) are included for comparison. $\Delta v_{p\text{-}p}$ of Si with $n_c = 7 \times 10^{19}$ cm$^{-3}$ and $n_c = 7 \times 10^{18}$ cm$^{-3}$ shows a $\sim T^{3/2}$ and $\sim T^{5/2}$ respectively near room temperature. (b) shows the residual linewidth of Si with $n_c = 7 \times 10^{19}$ cm$^{-3}$ (filled black circles) and $n_c = 7 \times 10^{18}$ cm$^{-3}$ (open black circles) after subtracting the phonon-broadened contribution predicted by Cheng et.al [21]. The residual linewidths show a temperature dependence slightly smaller than $T^{3/2}$.

We also investigated whether the integrated intensity of the EPR spectrum can be used for sensitive thermometry (using the spin-density contrast). In the non-metallic Si, the magnetic susceptibility of conduction electrons follows the Curie-Weiss law [42]. Therefore, we expect the temperature dependence of the integrated intensity to follow a $T^{-1}$ dependence. Fig.3 shows this is in fact the case in our measurements where at high temperatures, the integrated intensity of the EPR spectra in Si with $n_c = 7 \times 10^{16}$ cm$^{-3}$ follows a $T^{-1}$. The sensitivity of the integrated intensity to temperature in these samples is, therefore, -1 which is smaller compared to the temperature dependence of the linewidth. Even for the Si with $n_c = 7 \times 10^{18}$ cm$^{-3}$, the integrated intensity follows a $T^{-1}$ relationship.

In degenerately doped Si, we expect Pauli paramagnetism, where the susceptibility is independent of temperature. Our measurement on the silicon sample with $n_c = 7 \times 10^{19}$ cm$^{-3}$, shows no temperature dependence of the integrated intensity.

The difference of the temperature dependence of integrated intensity of extremely metallic and non-metallic Si also enables spin density contrast to decouple the contribution from concentration and temperature in systems with different doping levels.

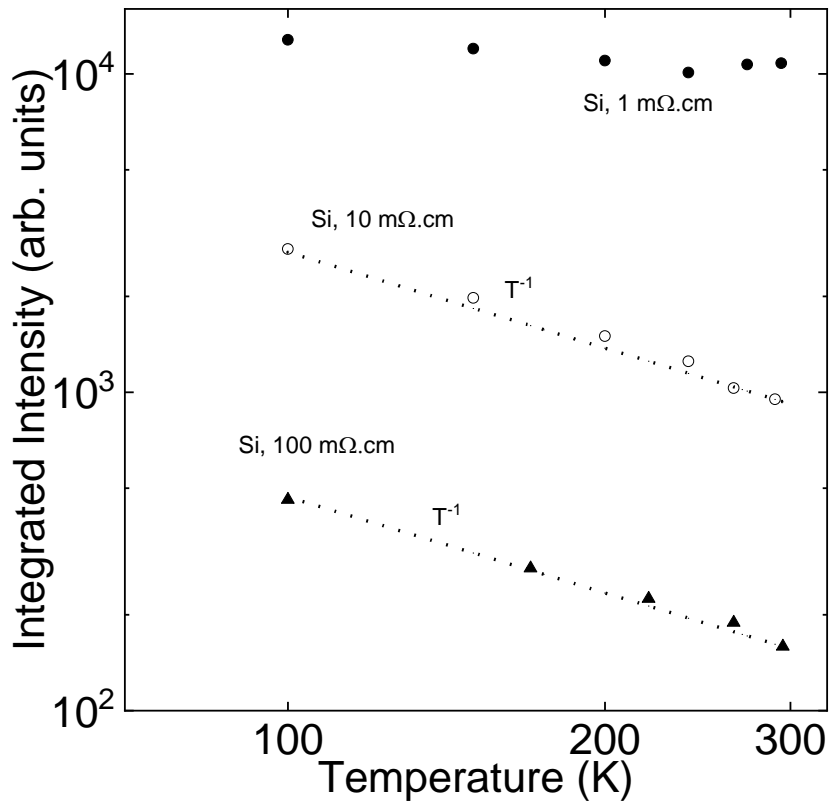

Fig 3. Temperature dependence of integrated EPR intensities for P-doped Si samples with carrier concentration $n_c = 7 \times 10^{19}$ cm$^{-3}$ (filled circles), $n_c = 7 \times 10^{18}$ cm$^{-3}$ (open circles) and $n_c = 7 \times 10^{16}$ cm$^{-3}$ (filled triangles). The presented data for each sample is normalized with Q-value of the resonance cavity at each temperature. The data for As-doped Ge is not included as the EPR intensity was too low to run EasySpin simulation and estimate the integrated intensity. P-doped Si samples with carrier densities $n_c = 7 \times 10^{18}$ cm$^{-3}$ and $n_c = 7 \times 10^{16}$ cm$^{-3}$ show a $T^{-1}$ dependence of the integrated intensity while the sample with $n_c = 7 \times 10^{19}$ cm$^{-3}$ does not show a significant temperature dependence.

Finally, we also investigated the temperature dependence of the linewidth and the integrated intensity of the states formed by ball milling on different silicon and germanium. For all Si samples these surface states gave a resonance at g ≈ 2.005 [33]. For Ge samples, the resonance of the surface state was at g ≈ 2.004 [33] . The integrated intensity of the surface states increased with the extent of milling and decreased with doping – we observed that small pieces of uncrushed wafers did not give any surface state resonance, while low doped Si showed resonance of surface states after crushing in mortar and pestle and heavily doped Si ($n_c = 7 \times 10^{18}$ cm$^{-3}$ and $n_c = 7 \times 10^{19}$ cm$^{-3}$) showed a surface state resonance only after ball milling. The surface states did not show any significant change in intensity and linewidth as a function of temperature. The results of the linewidth measurements of surface states are included in Supplemental Materials I.

Since the surface states do not offer any advantage in terms of thermometry, removing the surface states so that it does not affect the EPR image yields the best results for thermometry using semiconductors.

## IV. DISCUSSION AND CONCLUSIONS

We have systematically investigated the temperature dependence of linewidth and intensities of EPR of conduction electrons in Si at different doping levels and compared it to EPR of conduction electron in Ge. Our results show that while EPR in Ge and lightly doped Si ($n_c \leq 1 \times 10^{16}$ /cm$^3$) are not appropriate for thermometry near room temperature, heavily doped Si shows a strong temperature dependence of linewidth which could be used for 3D thermometry.

Since higher carrier densities and higher temperature sensitivity results in an increased SNR while broader linewidths mean reduced resonance lifetimes, we can define a figure of merit, $f$, for thermometry as the product of the carrier density $n_c$ and temperature sensitivity $S_T^{\Delta v}$ divided by the EPR linewidth $\Delta v$. Table 2 shows the figure of merit for Si with $n_c = 7 \times 10^{19}$ cm$^{-3}$, $7 \times 10^{18}$ cm$^{-3}$ and $7 \times 10^{16}$ cm$^{-3}$. It can be seen from Table 2 that increasing doping results in a better thermometer. Therefore, for thermometry applications where Si is embedded in a matrix, n-type Si with highest possible doping is optimal.

Table 2. Figure of merit, $f$, for EPR thermometry for silicon with $n_c = 7 \times 10^{16}$ cm$^{-3}$, $7 \times 10^{18}$ cm$^{-3}$ and $7 \times 10^{19}$ cm$^{-3}$. Carrier densities are normalized to $n_c = 7 \times 10^{16}$ cm$^{-3}$ in the table.

| Sample | Normalized carrier density ($n_c$) | Peak-peak linewidth ($\Delta v$) at 290 K (Gauss) | Sensitivity of linewidth to temperature: $S_T^{\Delta v} = \frac{d ln \Delta v}{d ln T}$ | $\frac{S_T^{\Delta v} n_c}{\Delta v}$ |
|---|---|---|---|---|
| C | 1 | 8 | 3 | 0.38 |
| B | 100 | 15 | 2.5 | 16.67 |
| A | 1000 | 30 | 1.5 | 50 |

A general problem for EPR imaging on heavily doped semiconductors is the reduced microwave penetration depth due to conductivity of the sample. Even for a microwave of frequency 1 GHz, the penetration depth for Si with resistivity 1 m$\Omega$.cm is only ~ 50 µm [39]. For imaging of devices with thick Si (e.g., 3D Integrated circuits), application of EPR for 3D thermometry is challenging.

To evaluate the possibility of embedding Si as a contrast agent in dielectric samples, we use effective medium theory to calculate the microwave penetration depth in the Si-dielectric composite. We have applied Bruggeman's effective medium theory [43] to evaluate the optical properties of a medium with water (resembling most tissues) and low volume fractions of Si.

The details of the effective medium theory and the values of the optical constants of Si and water are included in Supplemental Materials II.

Fig 4. shows the microwave penetration depth of water-Si composite for volume fraction of Si between 0-0.1 the volume of water at 1 GHz and 10 GHz for Si with carrier concentrations $n_c = 7 \times 10^{19}$ cm$^{-3}$, $7 \times 10^{18}$ cm$^{-3}$ and $7 \times 10^{16}$ cm$^{-3}$. Effective medium theory shows that, for very low volume fractions of Si in a host material, the microwave penetration depth is primarily determined by the microwave properties of the host material.

At 10 GHz, the penetration depth of microwave in water is less ~1 mm. Therefore, application of 3D thermometry in water using EPR is challenging. At 1 GHz, however, the penetration depth is ~ 10 cm, and 3D EPR imaging of systems with large volumes is possible. Therefore, our results show, embedding low volume fraction of heavily doped Si in biological tissues or other dielectric materials and performing measurements at frequencies (usually lower frequencies) where the absorption due to the material is low can be a possible means of sensitive 3D thermometry.

Imaging at low microwave frequency and low volume fractions, however, results in reduced signal-to-noise ratio. Future studies should include the quantification of the signal-to-noise ratio when small voxel sizes are imaged and lower frequencies are applied.

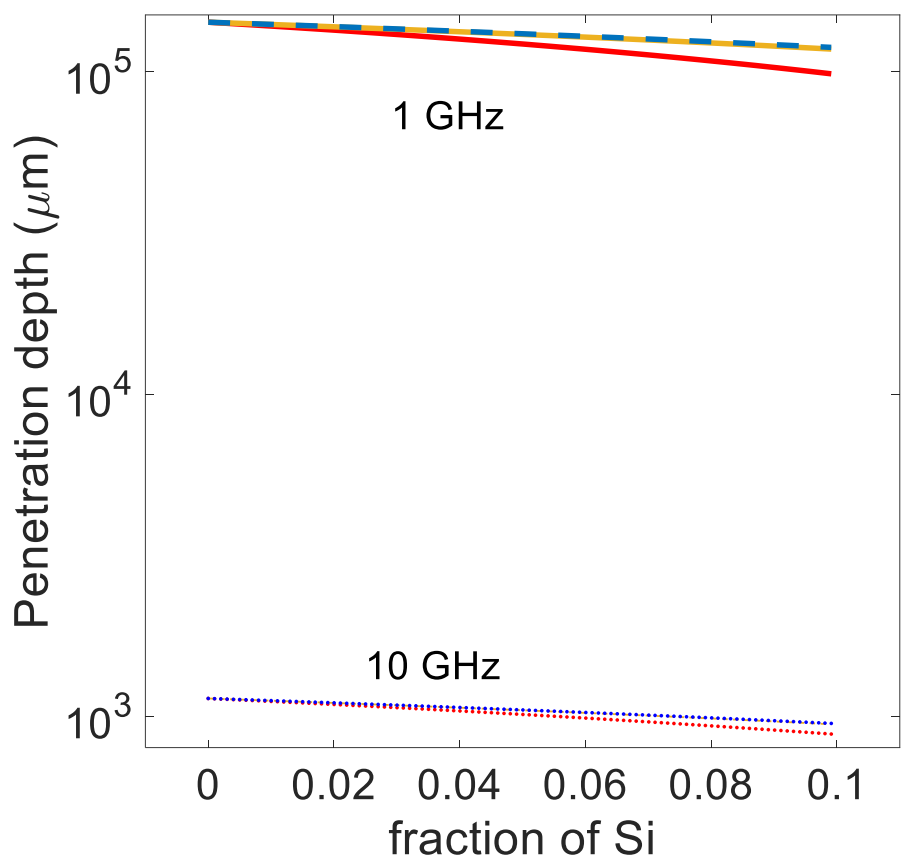

Fig 4. Calculated microwave penetration depth in a water– Si composite at 1 GHz (dotted lines) and 10 GHZ (red and orange solid lines and blue dashed line). The results are calculated using Bruggeman's symmetric effective medium theory for Si volume fractions from 0 to 0.1 times the volume of water and for Si carrier densities $n_c = 7 \times 10^{19}$ cm$^{-3}$ (blue), $7 \times 10^{18}$ cm$^{-3}$ (orange) and $7 \times 10^{16}$ cm$^{-3}$ (red).

**ACKNOWLEDGEMENTS**

The authors are grateful to Dr. Toby Woods of the School of Chemical Sciences, University of Illinois at Urbana Champaign for assistance with the EPR measurements. This project was funded by Semiconductor Research Corporation (Task ID: 3044.001). Major funding for Bruker EMXPlus